\documentclass[namedreferences]{solarphysics}
\usepackage[hyperref,optionalrh,showbiblabels]{spr-sola-addons}
\usepackage{graphicx}
\graphicspath{{figuras/}}
\usepackage{color}                       
\usepackage{breakurl}                         
\usepackage{booktabs}
\usepackage{multirow}
\usepackage[table,xcdraw]{xcolor}
\usepackage{float}
\usepackage{soul}

\chardef\us=`\_

\newcommand{\ds}[1]{{\color{red}#1}}

\begin{document}

\begin{article}

\begin{opening}

\title{Inferring the magnetic field asymmetry of solar flares from the degree of polarisation at millimetre wavelengths}


\author[addressref=aff1,email={douglas93f@gmail.com}]{Douglas F. da Silva}
\author[addressref={aff1,aff2}]{Paulo J. A. Sim{\~o}es}
\author[addressref=aff1]{R. F. Hidalgo Ram{\'i}rez}
\author[addressref=aff1]{Adriana V\'alio}

%
%


\address[id={aff1}]{Center for Radio Astronomy and Astrophysics Mackenzie (CRAAM), School of Engineering, Mackenzie Presbyterian University,  S{\~a}o Paulo, Brazil}
\address[id={aff2}]{SUPA School of Physics and Astronomy, University of Glasgow, G12 8QQ, UK}

\begin{abstract}

Polarisation measurements of solar flares at millimetre-waves were used to investigate the magnetic field configuration of the emitting sources. We analyse two solar flares (SOL2013-02-17 and SOL2013-11-05) observed by the POlarisation Emission of Millimetre Activity at the Sun (POEMAS) at 45 and 90 GHz, at microwaves from 1 - 15 GHz by the Radio Solar Telescope Network (RSTN), and at high frequencies (212 GHz) by the Solar Submillimetre Telescope (SST). Also, hard X-rays from these flares were simultaneously detected by the Reuven Ramaty High-Energy Solar Spectroscopic Imager (RHESSI). The flux and polarisation radio spectra were fit using a model that simulates gyrosynchrotron emission in a  spatially-varying 3D magnetic field loop structure. For the {modelling}, the magnetic loop geometry was fixed and the field strength was the only free parameter of the magnetic field. In addition, a uniform electron distribution was {assumed by} the model, with the number density of energetic electrons and the electron spectral index as free parameters. The fitted model reproduced reasonably well the observed degree of polarisation and radio flux spectra for each event yielding the physical parameters of the loop and flaring sources. Our results indicate that the high degree of polarisation during a solar flare can be explained by two sources located at the {footpoints} of highly asymmetric magnetic loops whereas low polarisation degrees arise from footpoint sources of symmetric magnetic loops.


\end{abstract}

%
\keywords{Flares, circular polarisation, millimetre wavelengths, magnetic fields}

\end{opening}

%
\section{Introduction}

Solar flares are rapid and intense brightness variations of solar emission at all bands of the electromagnetic spectrum. It is believed that the energy released during these events is of magnetic origin and occurs within active regions of the solar atmosphere, more likely at the top of magnetic loops. The energy released due to magnetic reconnection promotes {the} acceleration of particles, radiation, and heating of the plasma.  A fraction of these accelerated particles is injected into the magnetic loop towards the lower atmosphere \citep{fletcher:2011}, whereas the remaining particles travel in the opposite direction, towards the interplanetary medium, producing type III radio emission \citep{reid:2014}. The charged particles injected downwards into the magnetic loop exhibit a gyration depending on the direction of the magnetic field, producing polarised gyrosynchrotron radiation \citep{Ra_1969}. 

The microwave emission produced during a solar flare can be expressed as a linear combination of ordinary (O) and extraordinary (X) wave modes.  
{In the optically thin region, the polarisation direction corresponds to the X mode, whereas in the optically thick region to the ordinary mode  \citep{Ra_1969}.}
The polarisation in these two modes are opposite in direction and tend to be circular \citep{gary_book}.
Thus each polarity is associated with different legs of asymmetric magnetic loops. \citep{enome_1969,kundu1979,kundu2001ApJ...563..389K,kundu:2009,simoes:2013ApJ...777..152S}. 

Clues of this expected asymmetry of magnetic loops have also been obtained from X-ray observations. \cite{sakao_1994}, using data from Yohkoh's Hard X-ray Telescope (HXT), observed an asymmetry in intensity and size of hard X-ray (HXR) sources, with the source at a region of stronger magnetic field producing {fewer} X-rays, in about 75\% of his sample of events. Sakao's interpretation was that at the footpoint with weaker magnetic field there is greater electron precipitation. This is due to the smaller convergence of the field lines, which allowed more electrons to reach the chromosphere and thus generated more X-rays. Similar studies were performed by \cite{goff2004A&A...423..363G} and \cite{yang2012ApJ...756...42Y}. Their results are {in} agreement with Sakao's, with about 44\% and 75\% of their respective samples showing stronger HXR emission from regions with weaker magnetic field intensity, as measured by the Michelson Doppler Imager \citep[MDI, ][]{Scherrer2012SoPh..275..207S}. In the radio domain, \cite{kundu_1995} used 17 GHz data to investigate the existence of magnetic asymmetries in magnetic loops. The authors found that the source of intense hard X-rays was associated with unpolarised radio source, whereas the stronger source of radio emission was located over the region with stronger magnetic fields and was also polarised. 

Microwave observations from flares are currently one of the best methods to infer the coronal magnetic field in flaring regions, when done in conjunction with models for the gyrosynchrotron emission{ \citep{costa_rosal,Ra_1994,Stahli1989,Fleishman278}}. This method has also been successfully applied to estimate the magnetic field of coronal mass ejections  \citep{bain:2014ApJ...782...43B,carley:2017A&A...608A.137C}. The usual limitation of this method is that it provides only an effective intensity of the magnetic field of the emitting region and it is mostly imposed by the assumed geometry of the model. Several 2D and 3D loop models have been developed in the past decades, with increasing levels of spatial and spectral resolution {\citep{klein:1984A&A...141...67K,1984A&A...139..507A,simoes:2006A&A...453..729S,KuznetsovNitaFleishman:2011,Osborne:2019,kuroda_2018,fleishman_2018ApJ}} and applied to help in the interpretation of microwave flare data \citep[e.g.][]{simoes:2010SoPh..266..109S,MossessianFleishman:2012,reznikova2014ApJ...785...86R,nita2015ApJ...799..236N,MorgachevKuznetsovMelnikov:2015,gordo2017A&A...604A.116G,cuambe:2018MNRAS.477.1508C,gordo2019AdSpR..63.1453G}. 

{In the last decade, with the increase in computing power, improvements in the gyrosynchrotron emission algorithms, availability of routines for measurements of the photospheric magnetic vector, as well as codes for coronal magnetic field reconstruction, it is now possible to use 3D loop models in the calculation of the gyrosynchrotron emission from flares, and thus allow the investigation of  asymmetries in magnetic loops {\citep[e.g.][]{fleishman_2016ApJ}}}.
{
Despite several studies showing the importance of the optically thick part of the microwave spectrum \citep{bastian_1998,article,shevgaonkar}, source opacity and projection effects make the interpretation of optically thick emission difficult \citep[e.g.][]{simoes:2006A&A...453..729S}. Nevertheless, given the typical source conditions, the millimetric emission is  expected  to  be  optically  thin,  thus  providing  a  more  direct  view into the properties of the accelerated electrons and the magnetic field of the source.}

Observations of flare in the millimetric range are still less common than observations in the centimetric range. In the past 30 years, flare observations have been made with the Berkeley-Illinois-Maryland Array \citep[BIMA][]{kundu1991AdSpR..11...91K,Silva:1996ApJ...458L..49S,kundu2000ApJ...545.1084K}, the K\"oln Observatory for  Submillimeter and Millimeter Astronomy \citep[KOSMA, e.g.][]{luthi2004A&A...415.1123L} {,} and the Submillimeter Solar Telescope \citep[SST, ][]{Kaufmann1994,raulin:1999ApJ...522..547R,raulin:2004SoPh..223..181R,gimenez2009A&A...507..433G,gimenez:2013SoPh..284..541G}, with the latter still in operation. 

From November 2011 through the end of 2013, the POlarisation Emission of Millimetre Activity at the Sun \citep[POEMAS, ][]{Valio2013} telescopes observed the Sun at 45 and 90 GHz in both right and left circular polarisation, with a high temporal cadence of 10~ms. During this observing period\ds{,} several flares were detected; \citet{Hidalgo2019} investigated centre-to-limb effects on the flux density and polarisation of the millimetric emission, finding a correlation between the heliocentric position and the polarisation degree in a sample of 30 events. 

We present an analysis of two solar
flares (SOL2013-02-17T15:52 and
SOL2013-11-05T18:13) observed with POEMAS
to study their magnetic field
configuration, in particular, the
magnetic asymmetry of the footpoints of
the flaring loops. We employed different
observational datasets, described in
Section~\ref{sec:data}, to constrain many
of the free parameters of the model,
which was then used to fit the flaring
spectral data. The characteristics of the
magnetic field were inferred from a model
developed by \cite{Simoes_dr}, which
calculates the gyrosynchrotron radiation
in {a} spatially-varying magnetic field 3D
loop, described in
Section~\ref{sec:model}. Our strategy was
to use the observed polarisation degree
observed by POEMAS to restrict the model
fit to infer the magnetic field asymmetry during the flares. The results, discussion, and conclusions of this work are detailed in Sections{~\ref{sec:results}}, \ref{sec:discussion}, and \ref{sec:conclusion}, respectively.


\section{Observational Data}\label{sec:data}

In this work, we investigated two solar flares, SOL2013-02-17T15:52, an M1.9 GOES class event, and SOL2013-11-05T18:13, an M1.0 GOES class event. Both flares were observed at 45 and 90 GHz, in left- and right-handed circular polarisation by POEMAS. The calibration method to obtain the flux density at 45 and 90 GHz is described in \cite{Valio2013}. We also employed microwave data from the Radio Solar Telescope Network \citep[RSTN, ][]{guidice1979BAAS...11..311G}, from the Sagamore Hill Observatory. SOL2013-02-17 was also observed at 212 GHz by Solar Submillimeter Telescope \citep[SST, ][]{Kaufmann1994}.  

The combined radio observations from 5 to 212 GHz (RSTN, POEMAS, and SST) were used to build the radio flare spectra for our analysis. Using POEMAS's flux density of the left ($L$) and right ($R$) circular polarisation channels at 45 and 90 GHz we obtained the degree of polarisation, defined as:
\begin{equation}
    p_\mathrm = \frac{R - L}{R + L}.
    \label{eq:pol}
\end{equation}

Since these radio observations have no spatial resolution, to obtain information about the loop configuration and position where these sources are located we rely on ultraviolet (UV) images from the Atmospheric Imaging Assembly \citep[AIA, ][]{lemen_sdo}, on board of the Solar Dynamics Observatory \citep[SDO, ][]{pesnell2012SoPh..275....3P}. To locate the footpoints of the flaring loops, we employed the hard X-ray (HXR) imaging capabilities of the Reuven Ramaty High Energy Spectroscopic Imager \citep[RHESSI, ][]{lin_RHESSI}. RHESSI HXR images were constructed using CLEAN \citep{hurford_2002} with front detectors 2 to 8, with a beam width factor of 1.5 \citep[e.g.][]{SimoesKontar:2013}. All SDO/AIA and RHESSI images were taken near the peak time of the impulsive phase of the flares, {however} avoiding {saturated} AIA images.  Moreover, photospheric magnetograms from SDO's Helioseismic and Magnetic Imager \citep[HMI, ][]{Scherrer2012SoPh..275..207S} were used to estimate the highest magnetic field values at the flaring regions. All this information was used to constrain the magnetic field model of our 3D flaring loops.
 

\subsection{SOL2013-02-17}

\begin{figure}[t] 
   \centerline{\includegraphics[width=\textwidth]{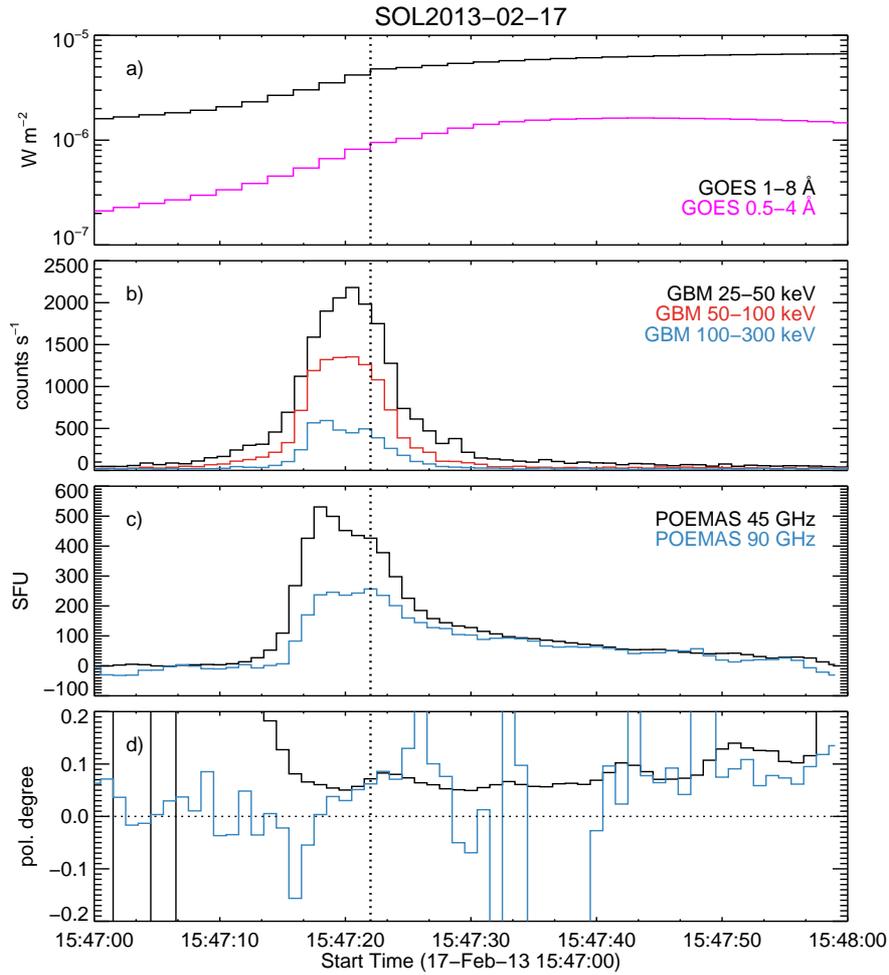}}
\caption{Lightcurves of SOL2013-02-17. (a) GOES SXR flux at 1--8 and 0.5--4 \AA, (b) HXR count rate from Fermi/GBM at 25--50, 50--100, 100-300 keV energy bands, (c) POEMAS 45 and 90 GHz flux density, and (d) polarisation degree. The vertical lines in all panels indicate the peak time of the 90 GHz emission, 15:47:22~UT, chosen for the analysis.}
\label{fig:lc17}
\end{figure}

SOL2013-02-17T15:52, a GOES class M1.9 event, occurred at the NOAA active region 11675 (N12E17). The time profiles of the event are shown in Figure~\ref{fig:lc17}: GOES soft X-ray channels 1--8 and 0.5--4~\AA\ (\ref{fig:lc17}a), Fermi Gamma-ray Burst Monitor \citep[GBP, ][]{megan_FERMI} HXR count rates at selected energy bands 25--50, 50--100, and 100-300 keV (\ref{fig:lc17}b), POEMAS 45 and 90 GHz flux (\ref{fig:lc17}c) and polarisation degree (\ref{fig:lc17}d). {The peak of the soft X-rays occurs later at 15:51 UT.}
The HXR and POEMAS lightcurves show a typical impulsive burst, occurring during the {rising} phase of the SXR emission. We defined the time for our analysis at the peak of the 90 GHz emission (15:47:22~UT, indicated by the vertical line in Figure \ref{fig:lc17}) to have a better signal-to-noise ratio for the spectra and polarisation data. The polarisation degree $p$ at 45 and 90 GHz (Figure~\ref{fig:lc17}d) indicates that this event is only weakly polarised, with $p$ below $+0.1$ (right-handed circular polarisation) at both frequencies, during the impulsive phase.

\begin{figure}[]   
\includegraphics[width=0.5\textwidth]{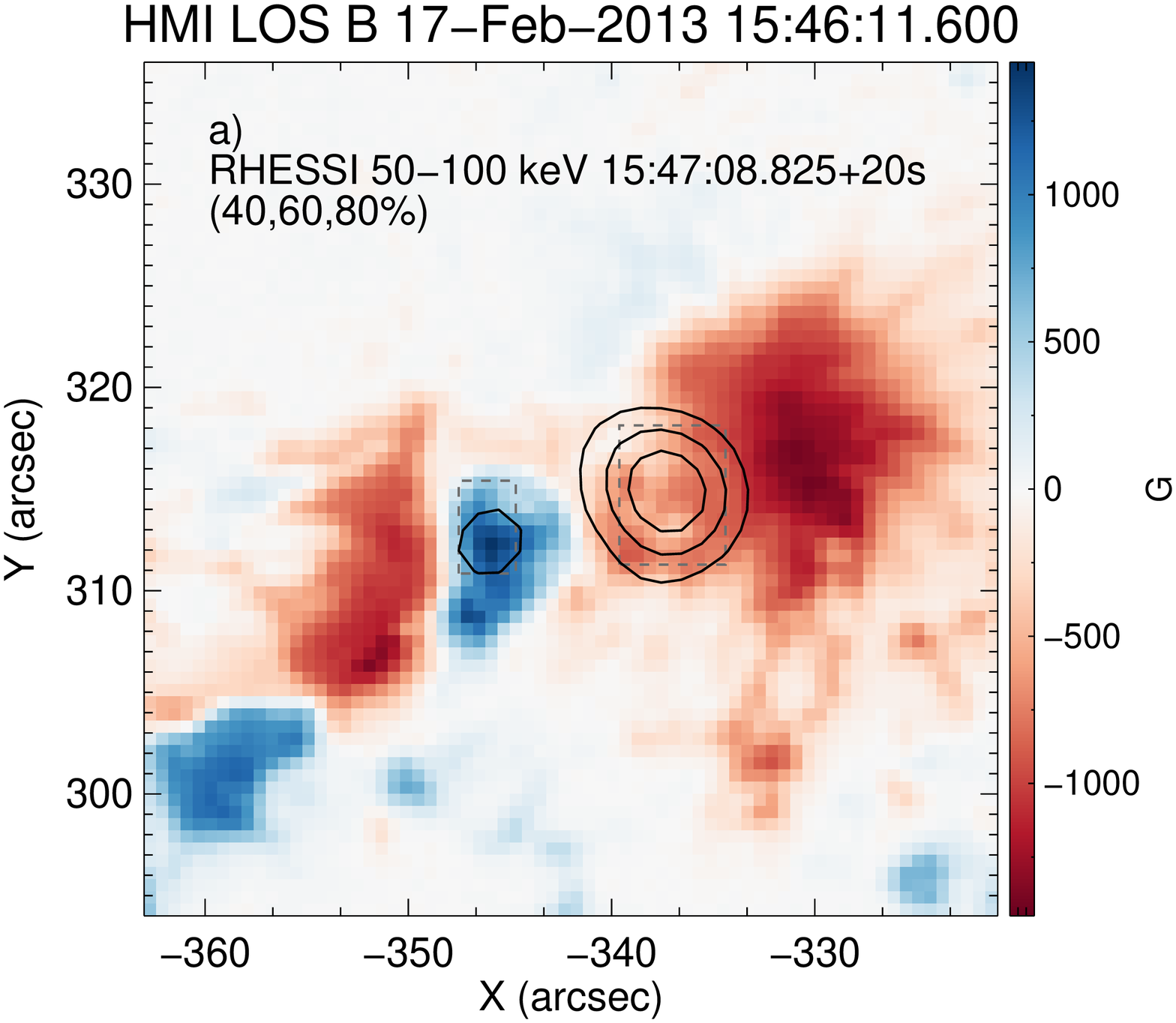}
\includegraphics[width=0.5\textwidth]{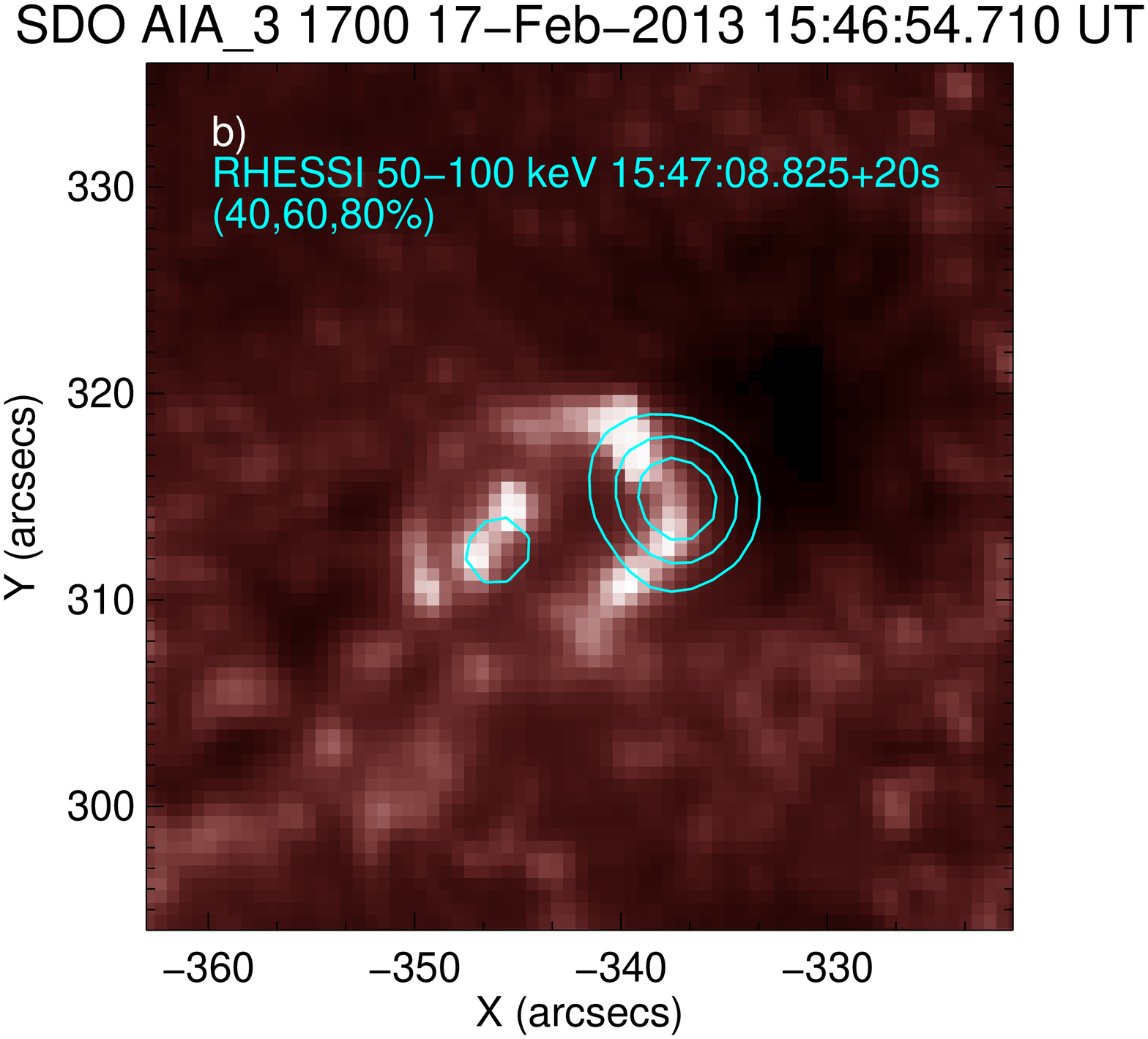}
\caption{
{(a)} {SDO/HMI line-of-sight (LOS) magnetogram of the flaring region taken near the peak time of the flare SOL2013-02-17T15:52, overlaid with RHESSI HXR 50--100 keV contours}.
{(b)} {SDO/AIA 1700 \AA\ image showing the bright flaring ribbons near the peak of the impulsive phase of the flare overlaid by RHESSI HXR 50--100 keV intensity contours at 40\%, 60\% and 80\%.}
}
\label{fig:map17}
\end{figure}

RHESSI HXR 50--100 keV intensity contours, highlighting the location of the main flare footpoints, have asymmetric intensities and both are well associated with the flaring ribbons seen in UV SDO/AIA 1700 \AA\ image (Figure~\ref{fig:map17}b). The UV ribbons and HXR footpoints are located at opposite sides of the magnetic neutral line as depicted by the SDO/HMI line-of-sight (LOS) magnetogram of the flaring region taken near the peak time of the flare (Figure~\ref{fig:map17}b).

\subsection{SOL2013-11-05}

\begin{figure}[] 
\centerline{\includegraphics[width=\textwidth]{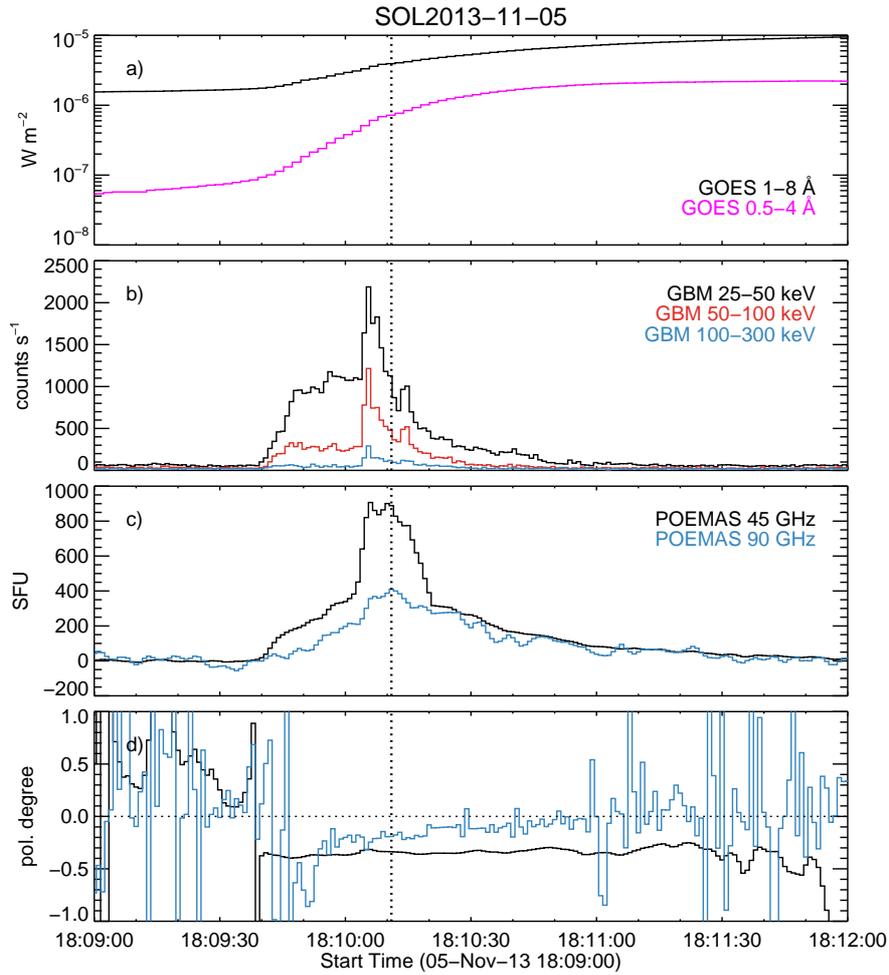}}
\caption{Lightcurves of SOL2013-11-05. (a) GOES SXR flux at 1--8 and 0.5--4 \AA, (b) HXR count rate from Fermi/GBM at 25--50, 50--100, 100-300 keV energy bands; (c) POEMAS 45 and 90 GHz flux density and (d) polarisation degree. The vertical lines in all panels indicate the peak time of the 90 GHz emission, 18:10:10~UT, chosen for the analysis.}
\label{fig:lc05}
\end{figure}

The flare SOL2013-11-05T18:13, a GOES class M1.0 event, occurred between 18:08:00 and 18:17:00 UT in NOAA active region 1890 (S12E47). Similarly to the previous event, the lightcurves for GOES SXR channels, Fermi/GBM HXR counts, POEMAS 45 and 90 GHz flux density and polarisation degree are shown in Figure~\ref{fig:lc05}. This event also has a typical impulsive behaviour, although it displays a slower rise in emission before the more impulsive peak seen at HXR and millimetric emission. The polarisation at 45 and 90 GHz {was predominant} in the $L$ sense, with values around $-0.4$ at 45 GHz and $-0.2$ at 90 GHz during the impulsive phase (see Figure~\ref{fig:lc05}d). 

\begin{figure}[]   
\centerline{
\includegraphics[width=0.5\textwidth]{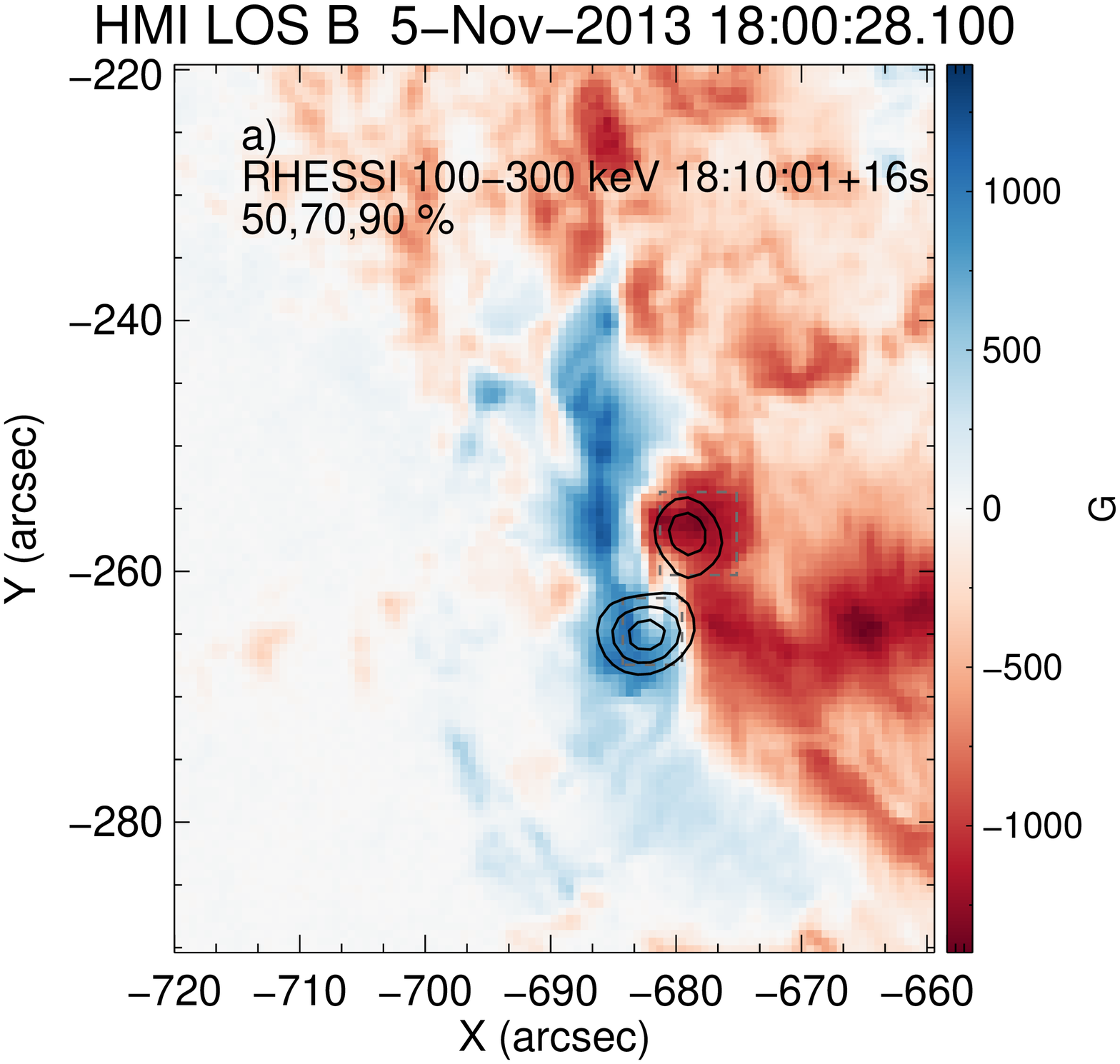}
\includegraphics[width=0.5\textwidth]{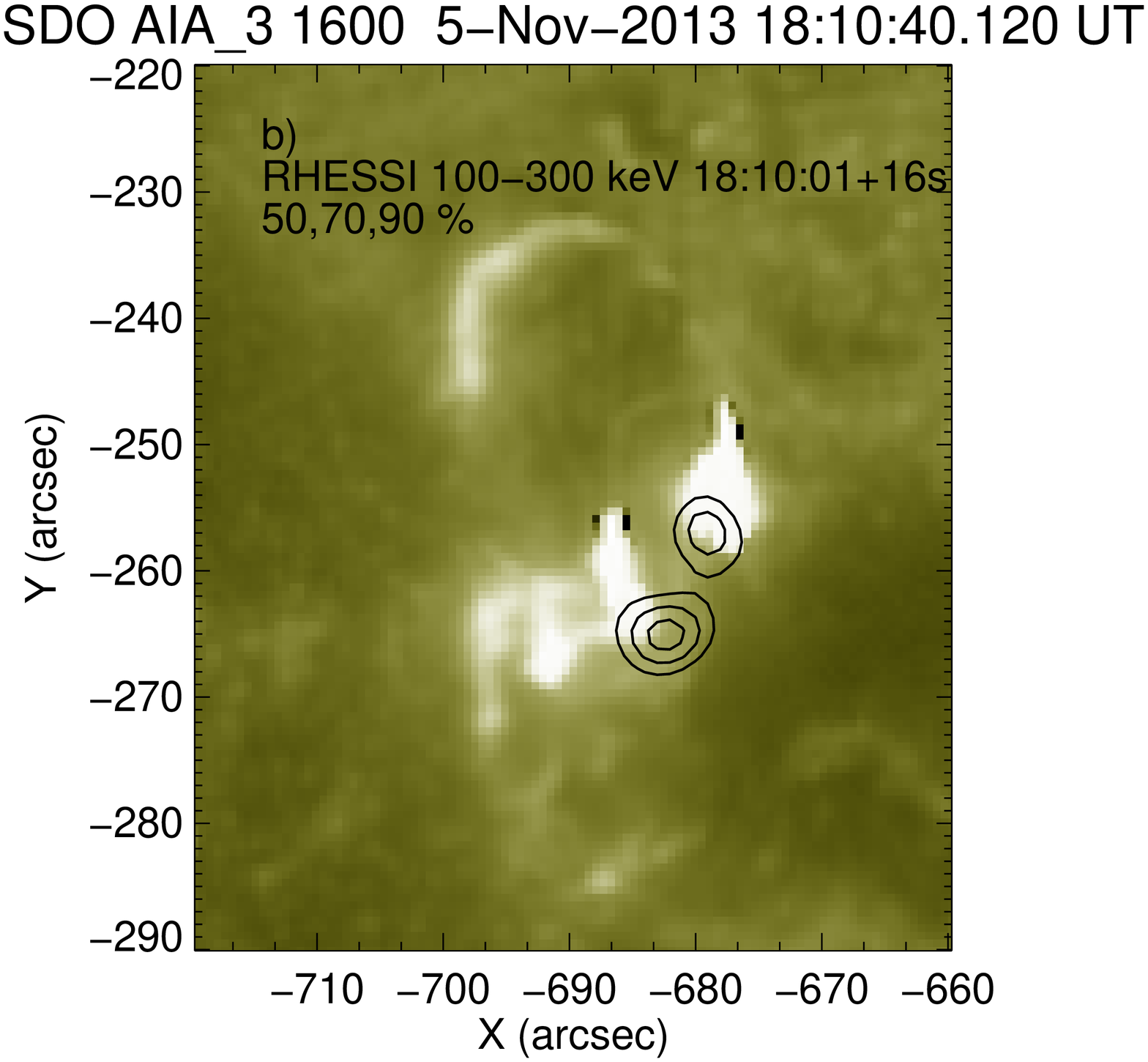}
}
\caption{
(a) SDO/HMI LOS magnetogram of the flaring region taken near the peak time of the flare SOL2013-11-05, overlaid with RHESSI HXR 50--100 keV contours.
(b) SDO/AIA 1600 \AA\ image showing the bright flaring ribbons near the peak of the impulsive phase of the flare overlaid by RHESSI HXR 100--300 keV intensity contours at 50\%, 70\% and 90\%.}
\label{fig:map05}
\end{figure}

The flare ribbons observed near the peak of the impulsive phase at 1600 {\AA} by SDO/AIA {are depicted in Figure~\ref{fig:map05}a}. The ribbons are well associated with the HXR 100--300 keV footpoints, indicated by the contours of the reconstructed RHESSI CLEAN image. The HXR footpoints and UV ribbons lie at opposite sides of the magnetic neutral line, revealed by the SDO/HMI LOS magnetic field {shown} in Figure~\ref{fig:map05}b.


\section{Description of the model}\label{sec:model}

\begin{figure}[]
\centerline{\includegraphics[width=0.55\textwidth]{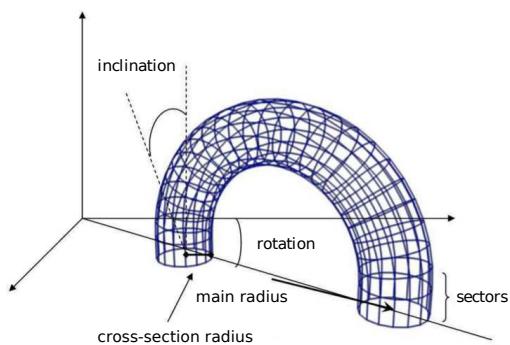}}
\caption{Geometry of the magnetic loop showing the parameters used in the model {\it gyro3D} \citep{Simoes_dr}}\label{fig:arco}
\end{figure}

We employed the model developed by \cite{Simoes_dr}, {\it gyro3D}\footnote{Available at \url{https://github.com/pjasimoes/gyro3d}} to construct a 3D 
loop model with an approximate geometry and location of each solar flare on the solar disk. {This model was then} fit the microwave spectrum and the polarisation degree at 45 and 90 GHz. The to simulates one single 3D semi-circular loop described by its length $L_{C}$ and cross-section radius $r$ to represent a flaring source. The orientation of the loop with respect to the observer is defined by its heliographic latitude $\Theta$ and longitude $\Phi$, rotation angle {relative} to the solar equator $\theta_r$, and inclination angle with respect to the local vertical $\theta_i$, as indicated in
Figure~\ref{fig:arco}. The loop is divided into $N$ uniform sections along its length. Each section is then associated with the parameters, as functions of the loop length $z$, to describe the magnetic field $B(z)$, temperature $T$ and density $n(z)$ of the thermal plasma and the distribution $f(E,\phi,z)$ of accelerated electrons (as a function of energy $E$ and pitch-angle $\phi$) necessary to calculate the gyrosynchrotron emissivity $j_\nu(z)$ and self-absorption $\kappa_\nu(z)$ coefficients. Here, the non-thermal electron population is assumed to have a power-law distribution in energy:

\begin{equation}
        f(\mathrm{E})= \mathrm{E}_\mathrm{min}^{\delta-1}~ (\delta-1)~\mathrm{N_{e}}~\mathrm{E}^{-\delta}
\end{equation}

with a spectral index $\delta$, normalised to the total electron density $N_e$ above a minimum energy cutoff $E_\mathrm{min}$. We also assume an isotropic pitch-angle distribution and uniform distribution of non-thermal electrons along the magnetic loop. 

The core program {\it
gyro}\footnote{Available at \url{https://github.com/pjasimoes/gyro}}
used to calculate these coefficients,
following \cite{Ra_1969}, is the same
program used by
\cite{cuambe:2018MNRAS.477.1508C}, and
also implemented by \cite{Osborne:2019}.
We refer to the latter for a complete
description of the computational
procedures used in the program to obtain
$j_\nu(z)$ and $\kappa_\nu(z)$. These
coefficients are obtained for both
ordinary and extraordinary wave modes, following \cite{Ra_1969}. The radiative transfer of the gyrosynchrotron radiation across the 3D source is calculated
following the same approach described in
\cite{simoes:2006A&A...453..729S}, where
each section of the loop is considered a homogeneous region, and therefore the integral of the specific intensity $I_\nu$ along the direction of propagation can be reduced to a sum. This calculation results in specific intensity $I_\nu$ maps at any chosen frequency, for the ordinary $I_+$ and extraordinary $I_-$ modes. The total flux density $F_\nu$ (in solar flux units, 1 sfu = $10^{-22}$~m$^{-2}$~Hz$^{-1}$) is obtained by summing the flux density of all pixels in the image, for both modes, taking into account the solid angle of the pixels.
{
The {gyrosynchrotron} theory is based on
moderately relativistic electrons  with
large Faraday rotation. In this case, the radiative transport equation decouples
into $\mathrm{I_{+}}$ and $\mathrm{I_{-}}$ \citep{Ra_1969}. The
polarisation degree $p$ obtained by
\cite{klein:1984A&A...141...67K} is given
by:
}
\begin{equation}
p=\mathrm{sign}( \cos \theta)\frac{I_+-I_-}{I_++I_-}
\end{equation}
{As the millimetre emission originates from moderately relativistic electrons \citep{kundu_1995}, this
equation is valid in this frequency
range.}
The calculations are performed from the reference frame of the electromagnetic waves, such that the extraordinary mode has right-handed polarisation and the ordinary mode, left-handed, for a positive $B$. As discussed in \cite{Osborne:2019}, in the reference frame of an observer, the polarisation is reversed, therefore $p_\mathrm{wave}=-p_\mathrm{obs}$. Our polarisation results are presented {in the} observer's frame.

To keep the number of free parameters for the model fitting to a minimum, we have fixed the following parameters, based on constraints from the {observational} data and simple assumptions: magnetic loop length, $L_{C}$, based on the HXR footpoint separation; inclination angle {relative} to the local vertical $\theta_i$; temperature $T$ and density $n(z)$ of the thermal plasma, since these parameters affect the free-free emission and Razin suppression effect, and are negligible in the millimetric range; and minimum and maximum energy cutoffs of the accelerated electron distribution, $E_\mathrm{min}$ and $E_\mathrm{max}$ respectively. These values of the fixed parameters are {listed} in Table ~\ref{tab:input}. Other parameters of the flares such as source heliographic position and loop size ($L_C$), that vary for each flare, were inferred from the UV and HXR images, and are also listed in Table~\ref{tab:input}. These parameters were the heliographic coordinates (longitude and latitude) and azimuth angle, or the viewing angle with respect to the solar equator. 

After fixing some of the input parameters, based on the observations or reasonable assumptions, we are left with the following free parameters: 
\begin{itemize}
    \item Magnetic field strength along the loop $B(z)$, defined by the magnetic field intensity at the top of the loop $B_0$ and the mirroring rate at each footpoint $\sigma$, according to the following quadratic expression \citep{kovalev_1976}:
        \begin{equation}
            B(z)= B_{0}\left [ 1+(\sigma-1)\frac{z^{2}}{L_{c}^{2}} \right ]
            \label{eq:Bloop}
        \end{equation}
    \item Spectral index $\delta$, and,
    \item Total electron density $N_e$.
\end{itemize}

Note that once the magnetic field at the looptop, $B_0$, and the mirror ratios $\sigma_1$ and $\sigma_2$ are {determined}, we can estimate the magnetic field at each footpoint: $B_+=\sigma_+ B_0$ and $B_-=\sigma_- B_0$ for the positive and negative footpoints of the magnetic loop{,} {respectively}.

To obtain the initial guess of the model parameters, we first use the {\it amoeba} function from Interactive Data Language (IDL), which performs the minimisation of the residual between observed data and model using the downhill simplex method.


As there is a high dependence on the magnetic field asymmetry on the result of the degree of polarisation it was necessary to obtain the uncertainty of this parameter. Thus the Markov Chain Monte Carlos (MCMC) method was used to achieve parameter uncertainties as well as to explore other parameter sets for minimisation. 
The minimisation of the free parameters was performed within the range shown in Table~\ref{tab:setting_values}. 
We propose a simplified model, with fixed  loop geometry, to fit the observed data, which assumes that the electrons {injected into the loop} {are} symmetric {and the transport} along the loop {is} uniform. Due to the application of this simplified model and its intrinsic limitations, the result of the fit to the observed polarisation and flux spectra is an approximation yielding order of magnitude values of the magnetic field and its asymmetry.

The upper limit for the magnetic field strength at the footpoints {was} inferred to restrict the upper boundaries of the loop top and the mirroring rates in the feet, {splayed} in Table ~\ref{tab:setting_values}{,} {were chosen so as to be compatible with the magnetic field observations}. 
The maximum absolute values of the line-of-sight magnetic field {at} the footpoints were obtained from HXR footpoints, indicated by dotted squares on Figures ~\ref{fig:map17}b and ~\ref{fig:map05}b. We obtained $-1972$ G and 1722 G for negative and positive magnetic regions respectively for SOL2013-11-05. For SOL2013-02-17, we found $-920$ G and 1270 G.  

\begin{table}[t]
\centering
\caption{Fixed input parameters to the model}
\label{tab:input}
\begin{tabular}{lcc}
\hline
Parameter      & SOL-2013-02-17     & SOL2013-11-05\\
\hline
Heliographic latitude $\Theta$ (deg) & 18 & -16 \\
Heliographic longitude $\Phi$ (deg) & -22 & -49 \\
Loop length ($L_{C}$) ($10^9$ cm) &  1     & 3  \\
Cross-section radius $r$ ($10^8$ cm) & 1.8 & 1.1 \\
Inclination angle $\theta{_i}$ (deg) & 0 & 0 \\
Rotation $\theta_r$ (deg) & 5 & -135 \\
Thermal electron density $n$ ($10^{10}$ cm$^{-3}$) & 1 & 1 \\
Thermal electron temperature $T$ (MK) & 10 & 10 \\
Minimum energy cutoff $E_\mathrm{min}$ (keV) & 10 & 10 \\
Maximum energy cutoff $E_\mathrm{max}$ (MeV) & 20 & 20 \\
\hline
\end{tabular}
\end{table}

\begin{table}[h]
\centering
\caption{Range of variation of free parameters for the model fit}.
\label{tab:setting_values}
\begin{tabular}{lcc}
\hline
Parameters  &  SOL2013-02-17 & SOL2013-11-05\\
\hline
Loop top magnetic field $B_0$ (G)  & $50 - 110$ & $50 - 145$\\ 
Mirror rate 1 $\sigma_1$& $1.0 - 14$ & $ 1.0 - 14$ \\
Mirror rate 2 $\sigma_2$   & $1.0 - 14$& $1.0 - 14$\\
Non-thermal electron density $N_e$ $10^7$ cm$^{-3}$ & $1 - 100$ & $(0.1 - 200)$ \\
Spectral index $\delta$& $1.0 - 6.0$ & $1.0 - 6.0$\\
\hline
\end{tabular}
\end{table}

The results from the calculation of the emission in a 3D magnetic loop yield the flux density and degree of polarisation spectra and location of the 45 and 90 GHz source. These results are listed in Table~\ref{tab:resultados}. Next, we describe the results for each event in detail.

\section{Results}\label{sec:results}

\begin{figure}[t]
\centering
{SOL2013-02-17}\par\medskip
\centerline{\includegraphics[width=0.85\textwidth,clip=]{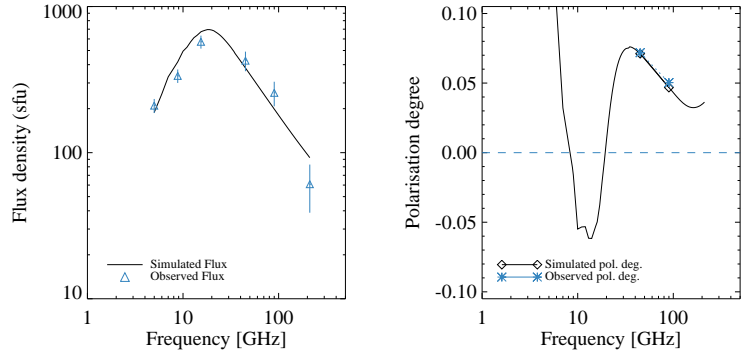}}
\caption{(a) Radio flux density spectrum at 15:47:22~UT, the peak of the 90~GHz emission of SOL2013-02-17. (b) 45 and 90 GHz polarisation degree spectrum at the same time interval. The observational data is shown as triangles and our model results shown by the solid line.}
\label{fig:spec17}
\end{figure}

\begin{figure}[t]   
\centering
{SOL2013-11-05}\par\medskip
\centerline{\includegraphics[width=0.85\textwidth,clip=]{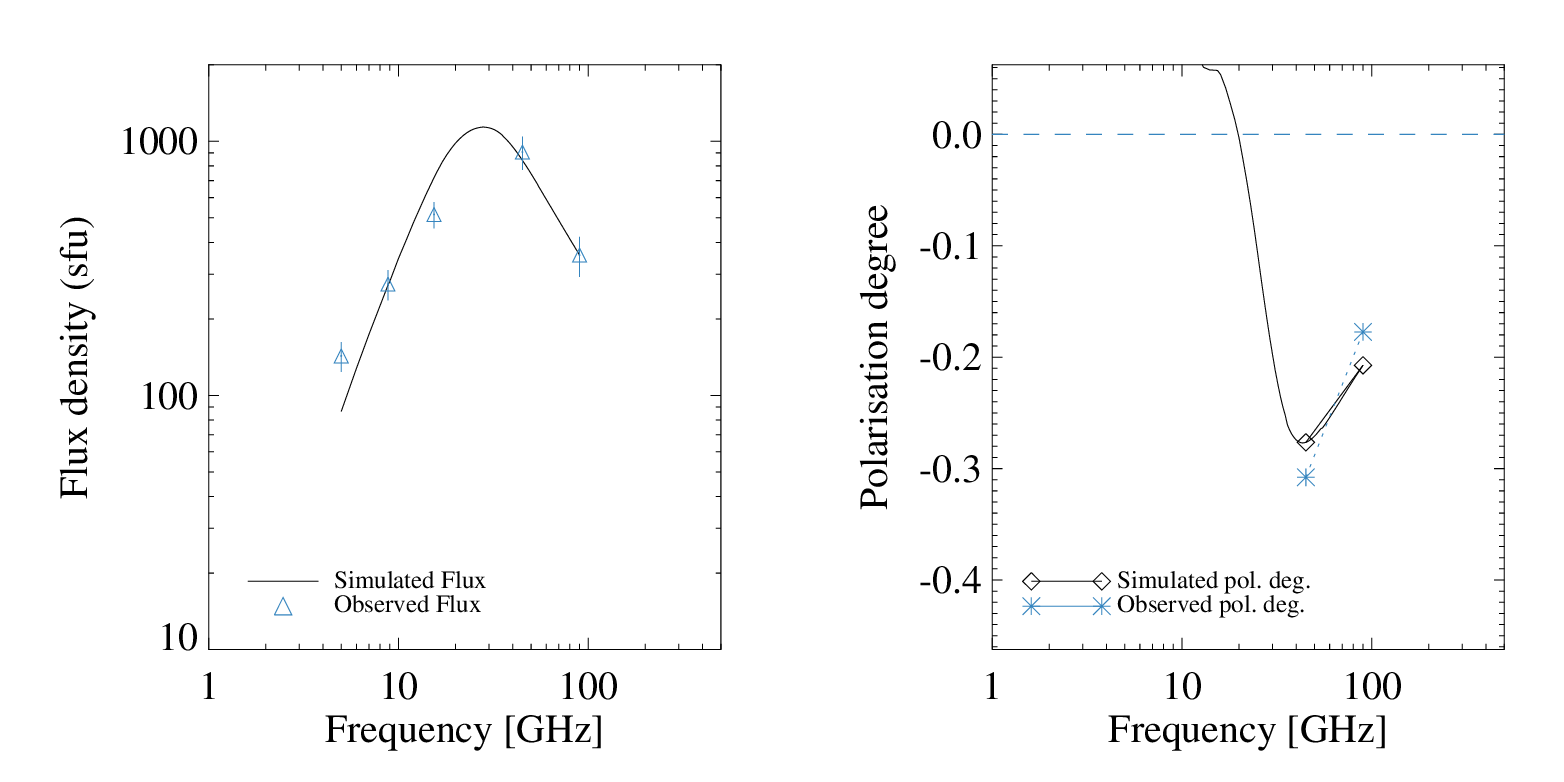}}
\caption{(a) Radio flux density spectrum at 18:10:10~UT, the peak of the 90 GHz emission of SOL2013-11-05. (b) 45 and 90 GHz polarisation degree spectrum at the same time interval. The observational data is shown as triangles and our model results shown by the solid line.}
\label{fig:spec05}
\end{figure}

The result of the model fit to the flux density spectrum and polarisation degree at the time of the peak 90~GHz emission, for SOL2013-02-17 and SOL2013-11-05 {are shown in Figures~\ref{fig:spec17} and ~\ref{fig:spec05}}, respectively. The observational data from RSTN, POEMAS, and SST are shown as triangles {whereas} the model {best fit} by the solid line. The parameters of the model fit are displayed in Table~\ref{tab:resultados}. The model fitting does not match all the data points, as expected, given the limitations and assumptions of our model. It can also be noted that the results of the magnetic field values of both events are lower than those obtained from the magnetogram where the events occurred. Smaller magnetic field values are expected as the {less than} 90 GHz {radio sources} comes {typically} from the {lower atmosphere}. 

\begin{figure}[]
    \centering
    \includegraphics[width=0.49\textwidth]{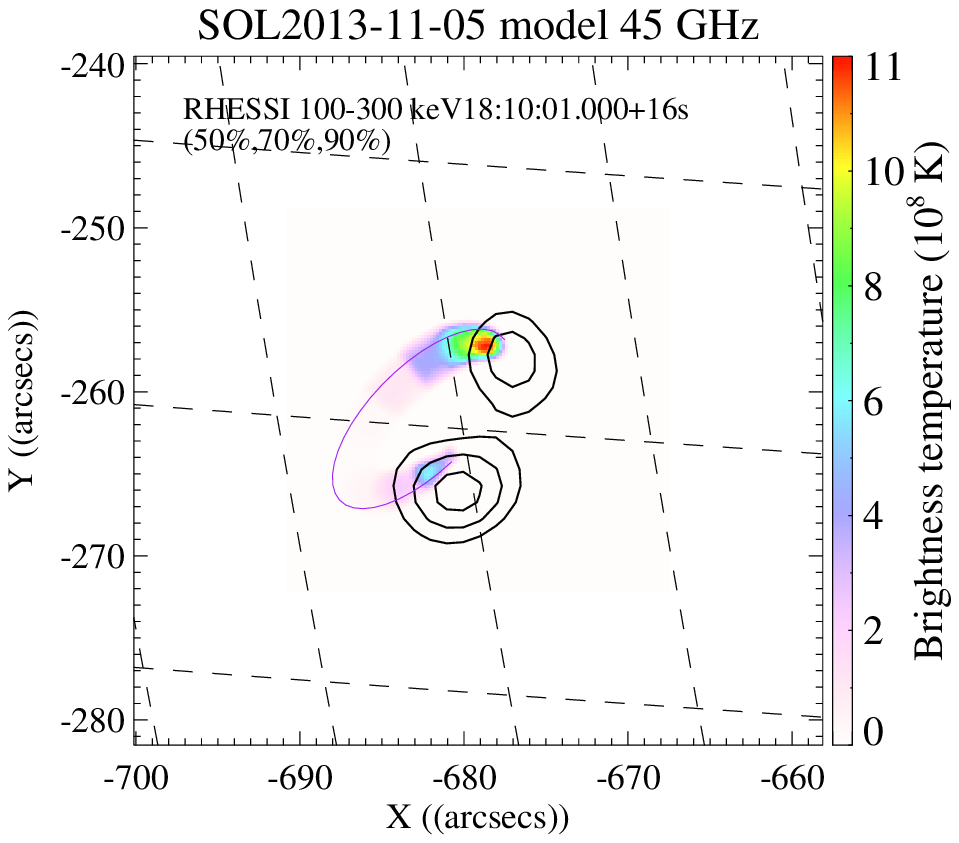}
    \includegraphics[width=0.49\textwidth]{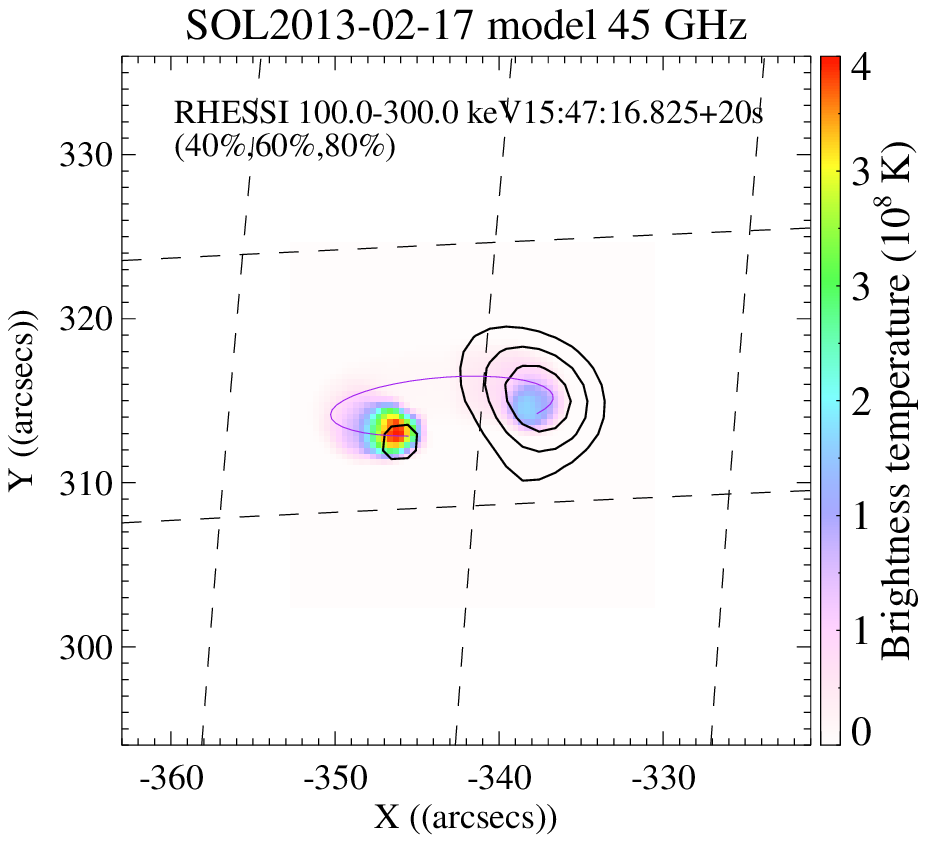}
    \caption{45~GHz brightness temperature maps from the model calculations for (a) SOL2013-02-17 and (b) SOL2013-11-05. RHESSI HXR contours are shown for reference. The solid purple line  shows the magnetic loop geometry of the model resulting from our model.}
    \label{fig:res}
\end{figure}

The calculated 45~GHz brightness temperature maps for SOL2013-02-17 and SOL2013-11-05 are shown in Figures ~\ref{fig:res}a and ~\ref{fig:res}b, respectively, overlaid with the corresponding RHESSI HXR contours (the same contours shown in Figures~\ref{fig:map17} and \ref{fig:map05}). The loop geometry of our model is {represented} by the purple line. 

As expected, the 45~GHz emission arises from the footpoints where the magnetic field is stronger. Since we assumed an uniform distribution of non-thermal electrons along the loop, the asymmetry in brightness is mostly due to the magnetic field intensity, that affects the amount of emission, and viewing angle that controls the observed polarisation. The calculated 90~GHz maps reveal the same brightness distribution as the 45~GHz maps{,} and thus are not shown here. 
Remembering that the fit results of both events were obtained for a fixed  magnetic field geometry, where the intensity was the only variable parameter of the magnetic field. 
Values of the resulting non-thermal electron density and the accelerated electron spectral index {despite being somewhat hard,} are typical of those found in the literature \citep[e.g.][]{bastian_1998}.

\begin{table}
\centering
\caption{Results from the model calculations.}
\label{tab:resultados}
\begin{tabular}{lcc}
\hline
Parameter & SOL2013-02-17& SOL2013-11-05 \\
\hline
Looptop magnetic field $B_0$ (G) & $82 \pm 2$ & $80 \pm 10$ \\
Footpoint magnetic field $B_-$ (G) & $-980 \pm 30$  &  $-1636 \pm 20$ \\
Footpoint magnetic field $B_+$ (G) & $+1020 \pm 40$ & $+860 \pm 40$ \\
Non-thermal electron density ($10^7 \mathrm{cm}^{-3}$) & $1.5 \pm 0.1$ &$47 \pm 1$ \\
Spectral index ($\delta$) & $2. 3\pm 0.1$ & $2.8 \pm 0.2$\\
\hline
\end{tabular}
\end{table}

\section{Discussion}\label{sec:discussion}



\cite{sakao_1994} investigated  the  asymmetry  of  HXR conjugated  footpoints. By using the magnetograms of the flaring regions, he associated the HXR asymmetry with the asymmetry of the magnetic field.
The more intense HXR emission originated where the intensity of the magnetic field was weaker, this was observed for 4 events out of 5 {analysed}. The explanation follows from the trap-plus-precipitation models \citep[e.g.][]{MelroseBrown:1976} in which footpoints with a weaker magnetic field {allow} more electrons to precipitate into the denser chromosphere, instead of being magnetically mirrored back to the coronal loops. By comparing HXR and microwave flare images, \cite{kundu_1995} addressed the same issue investigated by \cite{sakao_1994}. They found that at the sources with {a} stronger magnetic field, the radio emission is more intense whereas the HXR source is weaker.

Investigating 32 solar flares observed by Yohkoh, \cite{goff2004A&A...423..363G} found that 14 events follow Sakao's results. \cite{yang2012ApJ...756...42Y} {analysed} about 22 flares detected by RHESSI and found that 75\% are similar to that observed by \cite{sakao_1994}. Thus, previous authors generally confirm a consistent scenario with asymmetric magnetic mirroring.
%


Our results {for the} two flares demonstrated that the most intense {radio} source, from the fit at both frequencies (45 and 90 GHz), was located at the footpoint with the strongest magnetic field, which also coincided with the emission of the observed weakest hard X-rays source (see Figure~\ref{fig:res}).

\begin{figure}[]   
\centerline{\includegraphics[width=0.75\textwidth,clip=]{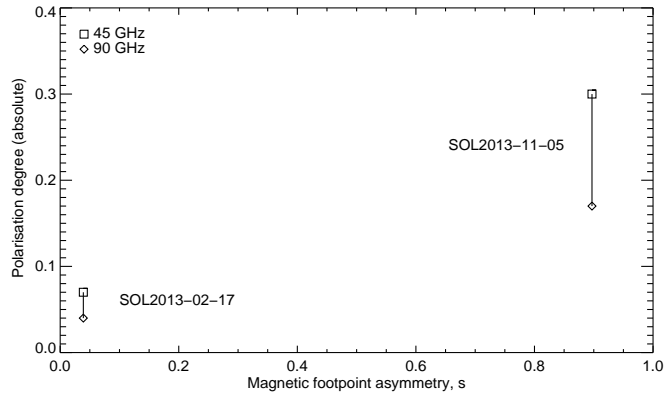}}
\caption{Polarisation degree as a function of the magnetic footpoint asymmetry, given by Eq.~\ref{eq:asym}.}
\label{fig:asy}
\end{figure}

To {analyse} the asymmetry of the magnetic loop, we define the {magnetic footpoint asymmetry}, $s$, given by 
\begin{equation}
s = {{\left | B_2 \right |-\left | B_1 \right |} \over \left | B_1 \right |}
\label{eq:asym}
\end{equation}
where $B_2$ is the stronger and $B_1$ the weaker magnetic field intensity of the footpoints. The {modeled} flaring loop of SOL2013-11-05 presented a footpoint magnetic field difference of about 50\%, and hence $s\approx 0.9$. For SOL2013-02-17, the intensity of the footpoints magnetic fields are very similar, giving {a} magnetic footpoint asymmetry of $s \approx 0.04$. In Figure~\ref{fig:asy} we show the modelled polarisation degree (absolute value) {versus} the asymmetry parameter of the magnetic field $s$. The highly polarised emission for SOL2013-11-05 ($p=0.2-0.3$) originated from a magnetic loop with a large magnetic field asymmetry (see Table~\ref{tab:resultados}). For SOL2013-02-17, the weaker polarisation degree ($p=0.04-0.07$) originated from an almost symmetric loop.

{
It follows, from these results, that it might be possible to infer the magnetic footpoint asymmetry even without spatial resolution. However, to confirm these results interferometric radio images are necessary}


However, we cannot rule out the effects of an asymmetric injection of non-thermal electrons in the loop system. Asymmetric reconnection \cite[e.g.][]{murphy2012ApJ...751...56M} can potentially lead to asymmetric outflows \citep[e.g.]{wang2017ApJ...847L...1W}, which may incur in different acceleration rates of electrons towards the conjugated footpoints. Moreover, the initial pitch-angle distribution of the accelerated electrons \citep[][]{LeeGary:2000} and/or asymmetric injection point \citep[][]{melnikov2002ApJ...580L.185M,rez2009ApJ...697..735R} are known to affect their spatial distribution along the magnetic loops and, as a consequence, will affect the polarisation of the observed emission. Although the main physical processes that control the transport of the non-thermal electrons are known \citep[e.g.][]{HamiltonLuPetrosian:1990}, we lack the sufficient and necessary spatial information about the geometry of the magnetic field in the corona that participates in flaring events. Along with the gyrosynchrotron calculations, electron transport models still have too many parameters that cannot be constrained by observational data, forcing any {modeling} efforts to rely on a large number of assumptions.
{However, 
interferometric images such as the ones produced by ALMA can greatly improve the diagnostics when combined with the spectrum and temporal variation of each source. Moreover, with spatially resolved images it is possible to vary the dynamics of the electrons during flares.}



\section{Conclusions}\label{sec:conclusion}

Here we {analysed} two flares observed at HXR and several radio wavelengths, with emphasis on the millimetric wave polarisation observed by the POEMAS telescopes. We investigated the solar flares SOL2013-02-17 and SOL2013-11-05, both observed by POEMAS with {a} good signal-to-noise ratio. 

Polarisation measurements by POEMAS at 45 and 90 GHz resulted in degrees of polarisation between 5\% and 40\% for the events. These events showed opposite sign for the degree of polarisation.

Keeping in mind the limitations of the model, such as a single magnetic loop and the uniform distribution of electrons on each {footpoint} of the magnetic loop, our main conclusion {is} that large polarisation {degree} at millimetre wavelengths from flares originates in highly asymmetric magnetic loops, {as seen in the event SOL2013-11-05}.

 Therefore, an asymmetric magnetic loop simulation in three dimensions is capable of reproducing the observed high polarisation degree of the emitting {sources} at 45 and 90 GHz. Hence to better understand solar flares, observations not only of the total flux density but also of the polarisation at several radio frequencies are necessary. Of course\ds{,} spatial resolution also is greatly desired.
 
 \begin{acks}
 We  acknowledge  partial  financial  support from FAPESP (grants $2009/50637-0$, 
$2013/24155-3$, $2013/10559-5$). P.J.A.S. acknowledges support from the University of Glasgow's Lord Kelvin Adam Smith Leadership Fellowship. The research leading to these results has received funding from the European Community's Seventh Framework Programme (FP7/2007-2013) under grant agreement no. 606862 (F-CHROMA). The authors are grateful to the late Prof. Pierre Kaufmann for the conception insight of the POEMAS telescopes. 
\end{acks}
%

%


%
%
\bibliographystyle{spr-mp-sola}
\bibliography{douglas_rb}  
%
%
%
%

\end{article} 
\end{document}